\documentclass[fleqn,10pt]{wlscirep}
\pdfoutput=1
\usepackage[utf8]{inputenc}
\usepackage[T1]{fontenc}
\usepackage{multirow}

\title{Tree size distribution as the stationary limit of an evolutionary master equation}

\author[1]{Szabolcs Kelemen}
\author[1]{M\'at\'e J\'ozsa}
\author[2]{Tibor Hartel}
\author[3]{Gy\"{o}rgy Cs\'oka}
\author[1,*]{Zolt\'an N\'eda}
\affil[1]{Physics Department, Babe\c{s}-Bolyai University, Cluj-Napoca, 400347, Romania}
\affil[2]{Faculty of Environmental Science and Engineering, Babe\c{s}-Bolyai University, Cluj-Napoca, 400294, Romania}
\affil[3]{Forest Research Institute, University of Sopron, M\'{a}traf\"{u}red, 3232, Hungary}

\affil[*]{zoltan.neda@ubbcluj.ro}

\begin{abstract}
The diameter distribution of a given species of deciduous trees in mature, temperate zone forests
is well approximated by a Gamma distribution. Here we give new experimental evidence for this conjecture
by analyzing deciduous tree size data in mature semi-natural forest and ancient, traditionally managed wood-pasture from Central Europe. These distribution functions collapse on a universal shape if the tree sizes are normalized to the mean value in the considered sample.
A novel evolutionary master equation is used to model the observed distribution. The model incorporates three probabilistic processes: tree growth, mortality and diversification. By using simple, and realistic state dependent kernel functions for the growth and reset rates together with an assumed multiplicative dilution due to diversification, the stationary solution of the master equation yields the experimentally observed Gamma distribution. The model as it is formulated allows analytically compact solution and has only two fitting parameters whose values are consistent with the experimental data for the growth and reset processes. Our results suggest also that tree size statistics can be used to infer woodland naturalness.
\end{abstract}
\begin{document}

\flushbottom
\maketitle
% * <john.hammersley@gmail.com> 2015-02-09T12:07:31.197Z:
%
%  Click the title above to edit the author information and abstract
%
\thispagestyle{empty}

\section*{Introduction}

The concept of universality in biological and social systems is highly debated~\cite{universality1,universality2,universality3,universality4,universality5,universality6,universality7}. Although many areas of science are keen to uncover universal statistical features of their studied systems, biology and sociology are usually focusing on quite the opposite, i.e. the specificities of the investigated problem. Beside this dominating trend, in ecology there are many  attempts for a unified statistical description of large plant or animal ensembles. Examples are population abundance studies~\cite{abundance1,abundance2}, scaling laws for size, life expectancy or motion trajectories~\cite{universality4,universality6}, topological features of food and metabolic networks and emerging patterns. 
In such a line of studies tree size evolution and the resulting statistics has been intensively studied in the past decades~\cite{Duncanson,gamma-iran}. Most of the models used in the literature are motivated by applications in sustainable forest management plans~\cite{forest-management}. 

Tree growth and mortality play a fundamental role in the ecosystem identity as well as the dynamics of forest and woodlands.
Exploring the potential universality of the dynamical mechanisms of tree ensembles (compact tree stocks) with different management and natural histories trough simple variables such as the tree size, remains an important statistical and modeling challenge. By validating the models and its assumptions on such statistical data one can then step further with the models and study the response of the system to environmental changes and human influence. Assuming argumentable growth and mortality rates, here we consider an analytically solvable evolutionary equation to model tree-size statistics in temperate zone woodlands.
 
Earlier statistical studies revealed that a Gamma distribution describes well tree diameter distribution in deciduous forest, although many other fitting functions were proposed~\cite{de_Lima_2015,gamma-iran,gamma-pdf-models}. Building on this finding, we employ a newly developed Local Growth and Global Reset (LGGR) model which is a simple evolutionary master equation with realistic dynamical assumptions~\cite{Biro-Neda,LGGR} to test the universality of tree diameter distribution using data from a wide range of forest and woodland ecosystems from Central Europe. Our data on individual tree diameters originates from temperate forests and woodlands covering a complete gradient of management history, from plantation forests (full human control), through semi-natural forests (reduced human interventions, multi-century continuity) till ancient wood-pastures with large old trees. In the following, first we will present the experimental data and then we will apply our model to analytically approximate the observed distributions and the real-life processes that are incorporated in the model.

\section*{Tree-size distribution revealed by the experiments}

Three different temperate zone woodland ecosystems types were selected for the tree-size measurements, with the aim of mapping various contributions to tree growth and mortality processes. 
We determined the mean Diameter at Breast Height (DBH) for all trees in compact well delimited regions for all the studied ecosystems. 

\begin{figure}[ht]
\centering
\includegraphics[width=\textwidth]{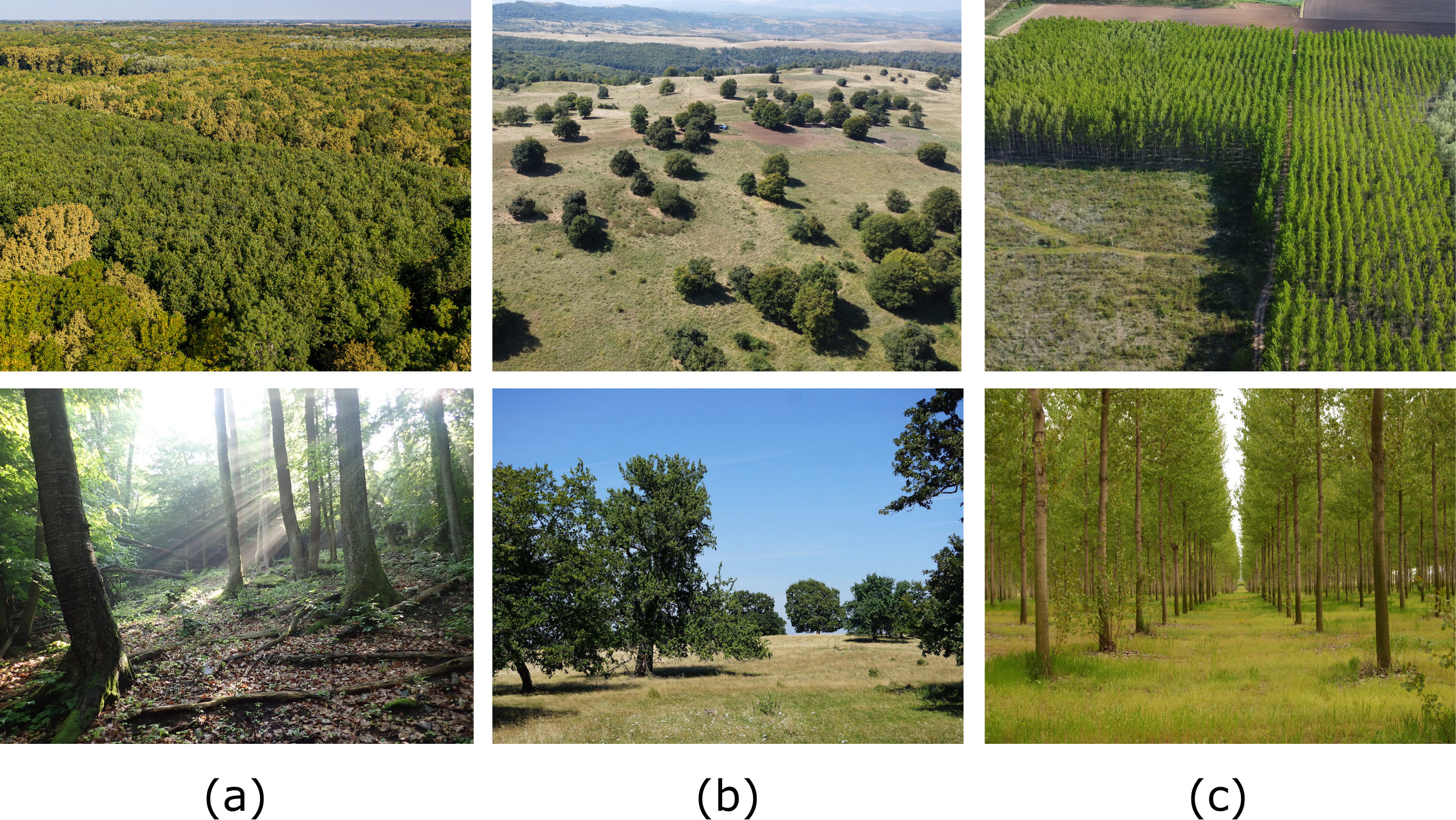}
\caption{Aerial (upper row) and ground level (bottom row) perspective image of the three ecosystems: semi-natural forest (a), natural wood-pasture (b), plantation (c). Source: Authors.}
\label{fig:figure1}
\end{figure}

Below we describe the three studied systems while the descriptive statistics of the trees are presented in Table~\ref{tab:data-all}. The first sample of trees originate from semi-natural, mature, deciduous forest (hereafter „forest”) from Central Romania (cca 400-600 m asl, Figure  \ref{fig:figure1}a). The dominant tree taxa which provides identity for these forests are the Oak ({\it Quercus sp.}), the Hornbeam ({\it Carpinus sp.}) and the Beech ({\it Fagus sp.}). The natural values of these forests are exceptionally high due to the low human interventions which allowed the accumulation of the dead wood and also the presence of large old trees~\cite{Dorresteijn_2013}. Forests from this region are covered by Natura 2000 protected area regulations. Grazing is prohibited in these forest since cca one century while the main economic use of the trees is the timber~\cite{Hartel_2015}. The density of trees is typically higher than 600 tree/hectare~\cite{Dorresteijn_2013}. The circumference of trees having at least 3 m height was measured at 130 cm from the ground~\cite{Hartel_2013}. Only the measurements from the dominant tree taxa (see above) were used in this study in order to increase the sample size. In order to avoid the forest edge effects on tree size the tree measurements were made at a distance of 270-850 m from the forest edge~\cite{Hartel_2013}. The age of the trees was estimated based on the rings counted in the field: from 15 to 250 years. Other, naturally established tree species which could present competition for the modelled trees are: {\it Acer pseudoplatanus, Acer platanoides, Tilia cordata} and in lesser extent {\it Prunus avium, Fraxinus excelsior} and {\it Acer campestre}.

The second sample of trees originates from an ancient, traditionally managed wood-pasture from Central Romania (cca 400-600 m asl,  Figure \ref{fig:figure1}b). The dominant tree taxa in the wood-pasture systems contains the three taxa mentioned above, and measurements of trees belonging to these taxa were used in this analysis. The origin of these wood-pastures is the centuries long silvopastoral use, when trees regenerated naturally, facilitated by thorny shrubs. Similarly to forests, the wood-pastures from this region are covered by Natura 2000 regulations. Unlike for the forests (see above), the main use of trees historically and now is the shade for livestock, fruits and errosion control for the soil~\cite{Hartel_2015}. The density of trees is around 7-25 trees/hectare~\cite{Hartel_2013}. The circumference of trees having at least 3 m hight was measured at 130 cm from the ground~\cite{Hartel_2013}.  The age of the trees based on ring counts ranges between cca 10 years to up to 300 years. Other, naturally established tree species: {\it Acer campestre, Pyrus sp., Malus sp., Prunus avium}~\cite{Hartel_2013}.

Finally, we considered hybrid Poplar tree ({\it Populus}) plantations with a density of approx. 400 trees/hectare, where all trees being planted in the same year and where no human intervention was considered since. These lately measurements were done in order to show the difference in the tree size-distribution for such controlled ecosystems, that did not reach a statistically stationary state, and mature natural forest environment with uncontrolled tree diversity and growth, where it is assumed that the tree-size distribution is stationary.  Another reason for studying such systems was to have information on the growth dynamics of genetically identical trees in controlled environments. The trees were planted in a regular square grid with an approximate distance of 5 meters between each other as it is illustrated in an aerial perspective in Figure \ref{fig:figure1}c. We made measurements for two plantations of different ages (approximately 10 and 15 years). Since virtually no other tree species was present in the plantations, we assume no interspecific competition in this system.

\begin{table}[ht]
\centering
\begin{tabular}{|c|c|c|c|c|c|}
\hline
{\bf Woodland type}   & {\bf Species / Stand age} & {\bf Nr. of trees} & {\bf Lowest DBH} [cm] & {\bf Greatest DBH} [cm] & {\bf $\langle$ DBH $\rangle$} [cm] \\ \hline
\multirow{3}{*}{Semi-natural forest} & Oak & 883 & 3.2 & 122.5 & 38.1 \\ 
\cline{2-6}
                            & Beech & 1782 & 3.2 & 115.2 & 31.5 \\
\cline{2-6}
                            & Hornbeam & 1994 & 1.6 & 76.4 & 20.1 \\ \hline
\multirow{3}{*}{Wood-pasture} & Oak & 1013 & 4.1 & 248.3 & 87.0 \\
\cline{2-6}
                            & Beech & 100 & 10.2 & 136.9 & 74.1 \\
\cline{2-6}
                            & Hornbeam & 255 & 4.8 & 202.1 & 54.0 \\ \hline
\multirow{2}{*}{Poplar plantation} & $\simeq 10 years$ & 1076 & 5.7 & 36.0 & 18.3 \\
\cline{2-6}
                            & $\simeq 15 years$ & 1613 & 5.1 & 54.7 & 27.5 \\ \hline
\end{tabular}
\caption{Statistical overview of the processed semi-natural woodland and plantation data.}
\label{tab:data-all}
\end{table}

\begin{figure}[ht]
\centering
\includegraphics[width=\textwidth]{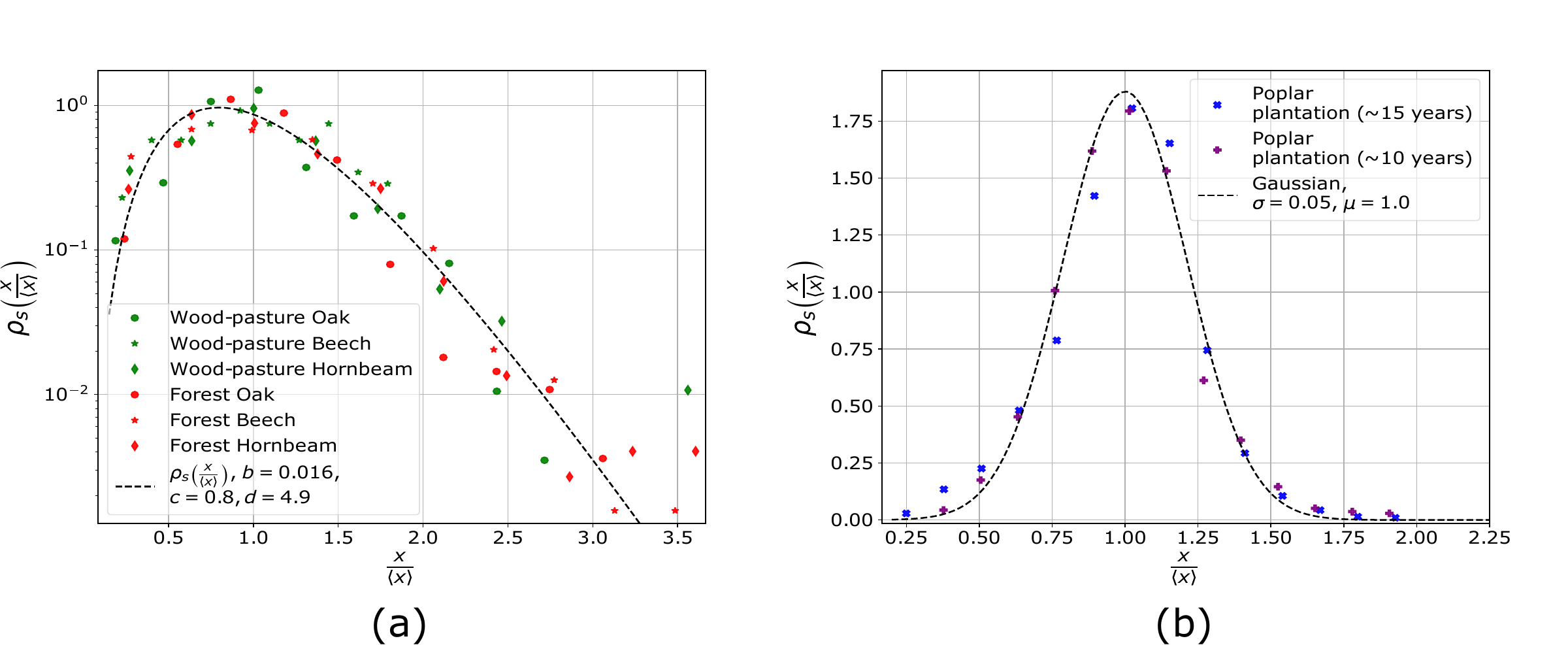}
\caption{Experimental DBH distributions (probability density functions)  (a) from natural forests (red), wood-pastures (green) and the Gamma fit obtained by the LGGR model (Equation \ref{eq:density-function-final}), (b) from poplar tree plantations and the Gaussian fit.}
\label{fig:distribs-exp}
\end{figure}

All three databases constructed by us contain exhaustive measurements in a compact tree ensambles for DBH values~\cite{data:DBH:data}. From the collected data we constructed the normalized probability density function for the tree size distribution. Tree sizes, $x$, are quantified with their DBH values and in our statistics these were normalized to the mean for the specific tree ensemble: $x \rightarrow y=\frac{x}{<x>}$. The $\rho(y)$ probability densities computed from the data are shown in Figure~\ref{fig:distribs-exp}. 

The tree size distributions for semi-natural forests and wood-pastures collapse on a master trend which can be well approximated with a Gamma distribution. Our finding on the goodness of the Gamma distribution is in agreement with earlier studies on tree-size distribution in forest environment~\cite{gamma-iran,gamma-pdf-models}. 

As expected, the statistics for the plantation is strikingly different, resembling a Gaussian trend (Figure \ref{fig:distribs-exp}). However, the distributions in $y$ for two different aged poplar plantations collapses again (Figure \ref{fig:distribs-exp}). The Gaussian nature of the distribution in a tree plantation seems consistent with what one would expect from simple analogies with similar statistics in other controlled biological systems. 

The Gamma type  tree-size distribution in a mature, semi-natural forest is however a more complex problem, and in understanding it one should follow the dynamical evolution of the tree ensemble, the interplay of growth and mortality processes. Due to their mature nature, one can then assume that the observed distributions are stationary ones, so the stationary limit of such an evolutionary equation should describe the observed distributions, which is a helpful assumption for modeling purposes. In the followings we will look deeper in the available statistical data on such systems and try to understand them through mean-field like evolutionary models.

\section*{The LGGR modeling framework}
\label{sec:LGGR-general}

For modeling purposes we used the Local Growth and Global Reset (LGGR) master-equation framework. This evolutionary type equation is a mean-field like description of an ensemble where individuals are subject to the same probabilistic local growth and global reset processes~\cite{LGGR}.
Reset is a process where an individual with a given state leaves the considered ensamble (either by mortality or some equivalent process) and it is replaced by a different individual in the ground state. For a unidirectional growth process this reset is needed in order, to achieve a stationary state.
 It has been proven to be appropriate for explaining various distributions that are characteristic for different complex systems~\cite{LGGR, ZNeda1, ZNeda2}. For illustrating such a dynamics let us consider that the states of the elements are characterized by a quantity $x$, in our case this quantity can be the size quantified by DBH.

In a first approach let us discretize the trees diameter in well distinguishable states, described by an integer number of corresponding DBH quanta, $n$ ($x\rightarrow n$). In this discrete scenario we denote by $N_n(t)$ the number of elements in state $n$ at time $t$. Assuming local probabilistic changes for the states of the elements and a probabilistic resetting process to the $n=0$ state, an evolutionary master-equation can be considered:
\begin{equation}
\frac{dN_n(t)}{dt}=\mu_{n-1} N_{n-1}+\lambda_{n+1}N_{n+1}-(\mu_n+\lambda_n+\gamma_n)N_n(t)+ N_{total} \delta_{n,0}\langle \gamma \rangle(t).
\label{equation:master_gen_N}
\end{equation}
Here $\mu_n$ is the state dependent local growth rate (probability per unit time) of going from state $n$ to state $n+1$, $\lambda_n$ is the local decrease rate of going from state $n$ to state $n-1$, and $\gamma_n$ is the reset rate for going from state $n$ to state $0$. 
The system preserves the $N_{total}=\sum_{i} N_i$  elements in the system by the last term, which is nonzero for $n=0$. We have thus:
\begin{equation}
\langle \gamma \rangle(t) =\sum_{j} \gamma_j \frac{N_j(t)}{N_{total}}.
\label{equation:mean_res_N}
\end{equation}
For many real-world processes, like the case of trees, the local dynamics is unidirectional. The living tree's diameter can only increase, with state dependent growth rates. This means that in Equation \ref{equation:master_gen_N} $\lambda_n=0$ for all $n$ states and the process becomes the one we named Local Growth and Global Reset (LGGR) dynamics:
\begin{equation}
\frac{dN_n(t)}{dt}=\mu_{n-1} N_{n-1}-(\mu_n+\gamma_n)N_n(t)+ N_{total} \delta_{n,0}\langle \gamma \rangle(t).
\label{equation:master_local_N}
\end{equation}
We can switch now the description from the $N_n$ occupancy numbers to the $P_n=N_n/N_{total}$ probabilities that a tree's DBH is $n$ quanta at time moment $t$. Naturally, normalization of $P_n(t)$ satisfies: $\sum_{\{n\}} P_n(t)=1$.  
The evolutionary master equation describing the local unidirectional transitions and a random resetting process is also a system of coupled first order differential equations:

\begin{equation}
\frac{dP_n(t)}{dt}=\mu_{n-1}P_{n-1}(t)-\mu_n P_n(t)-\gamma_nP_n(t) + \delta_{n,0}\langle \gamma \rangle(t) . 
\label{equation:master_dis}
\end{equation} 
\noindent
The last term in Equation~\ref{equation:master_dis} maintaining the normalization of $P_n(t)$ is :
\begin{equation}
\langle \gamma \rangle(t) =\sum_{j} \gamma_j P_j(t).
\label{equation:mean_res}
\end{equation}

\begin{figure}[ht]
\centering
\includegraphics[width=\textwidth]{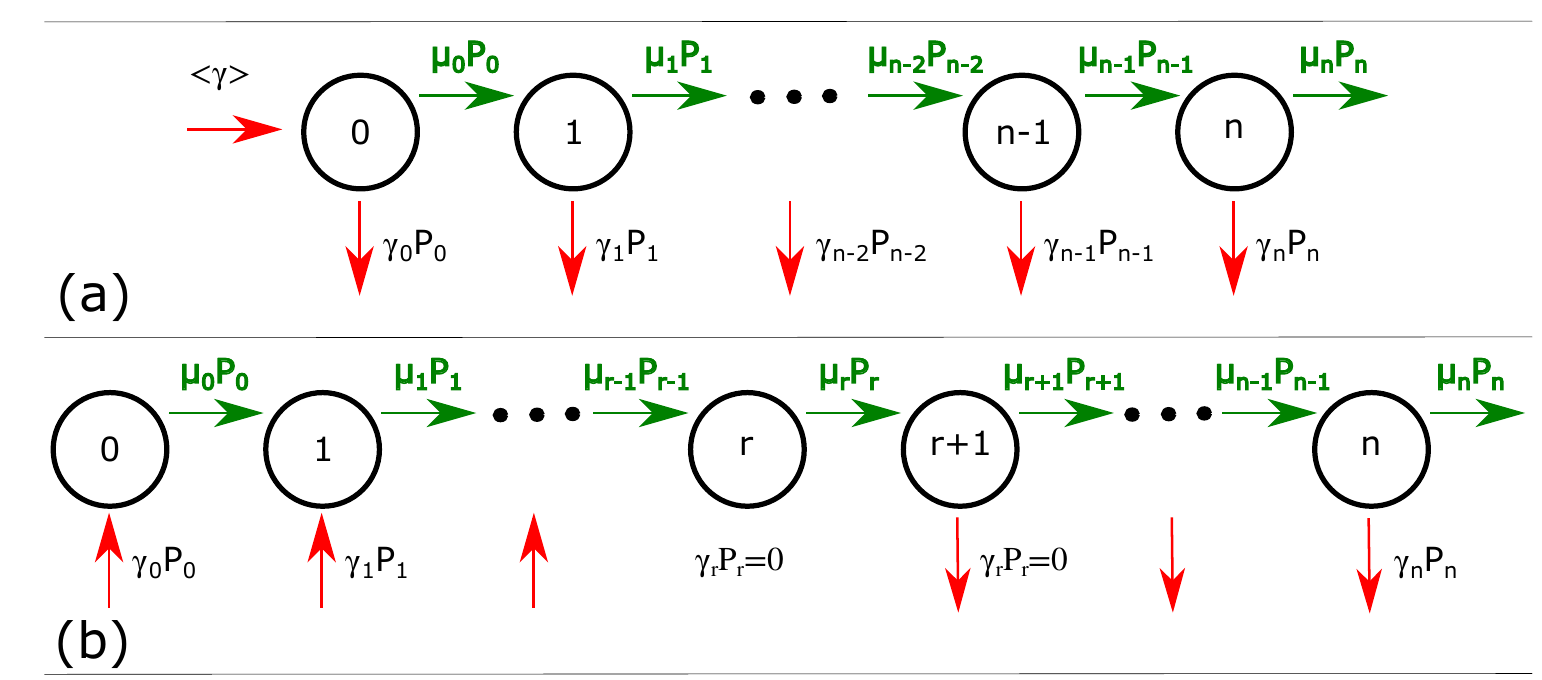}
\caption{Schematic illustration of the growth and reset process for two scenarios based on the form of the reset rate: (a) simple mechanism with only positive reset rate, (b) the reset rate can be both negative and positive ($\gamma_n<0$ if $n<n_r$, and  $\gamma_n>0$ for $n>n_r$).}
\label{fig:lggr_schem}
\end{figure}

\noindent
Based on the mathematical form of the reset rate, $\gamma_n$, two different dynamical scenarios can be distinguished. The simplest case is when for all $n$ values the state dependent reset rate, $\gamma_n$, is positive. Reset means that the element disappears from state $n$ and reapers in state $0$. For trees this simple reset describes tree mortality, and consequently the replacement of a tree with a new individual with $0$ size. This dynamics is represented in Figure~\ref{fig:lggr_schem}a. A more complicated dynamical scenario is when the reset rate, $\gamma_n$, can be both positive and negative as a function of the $n$ value. A scenario of this type is represented in Figure~\ref{fig:lggr_schem}b. One should keep in mind that a negative reset is an inverse process to the ordinary reset, it means that an element is appearing in state $n$ and disappears from another state, preserving the total balance. In the case of tree ecosystems this would mean that a new tree that appears in our statistics is characterized not by a $0$ size, but it appears in the $n>0$ bin, usually $n$ smaller than a critical $n_r$ value. Simultaneously, large trees are dying out or get harvested so they disappear from states with  $n>n_r$. This second scenario considering a state dependent smart reset rate offers much more flexibility and it is more appropriate for modeling the tree growth dynamics in the ecosystems where our data was collected from. 
Such an attempt was considered recently for modeling the distribution of wealth and income in human societies~\cite{ZNeda2,ZNeda3}. 

\noindent

Another possibility to include additional terms in the evolutionary equation is by considering the case when the number of elements are also changing in the system. For example, in case the number of elements are increasing 
(or decreasing) multiplicatively
\begin{equation}
\frac{dN_{total}}{dt}=\kappa_0 N_{total}(t),
\end{equation}
one gets
\begin{equation}
\frac{dN_n(t)}{dt}=\frac{d\, (N_{total}(t) P_n(t))}{dt}=N_{total}(t)\frac{dP_n(t)}{dt}+P_n(t)\frac{dN_{total}(t)}{dt}=N_{total}(t)\frac{dP_n(t)}{dt}+\kappa_0 N_{total}(t){P_n(t)},
\end{equation}
leading to and extra reset-type term in the master equation for $P_n(t)$:
\begin{equation}
\frac{dP_n(t)}{dt}=\mu_{n-1}P_{n-1}(t)-\mu_n P_n(t)-(\gamma_n+\kappa_0)P_n(t) + \delta_{n,0}\langle \gamma \rangle(t).
\label{equation:master-discrete}
 \end{equation}
Such kind of process was recently considered  for explaining the universal statistics of citations and Facebook shares~\cite{ZNedaFacebook}.

Handling mathematically the coupled differential equations from Equation~\ref{equation:master-discrete} in the discrete dynamical picture is quite tedious. The discrete process described by Equation~\ref{equation:master-discrete} can be generalized to continuous states ($n \rightarrow x$) in the limit $dt \rightarrow 0$~\cite{Biro-Neda}. In such a picture, instead of the discrete state probabilities $P_n(t)$ we will have the continuous probability densities $\rho(x,t)$ with the normalization condition $\int_{\{x\}} \rho(x,t) dx=1$. The growth and reset rates are written as functions of the state variable $x$:
\begin{equation}
 \begin{aligned}
 \mu_{n} \rightarrow \mu(x) \\ \gamma_n \rightarrow \gamma(x) \\ \kappa_0 \rightarrow \kappa.
 \label{reset_growth}
 \end{aligned}
 \end{equation}

\noindent
By taking this continuous state generalization, the master equation written in Equation~\ref{equation:master_dis} transforms into a partial differential equation:

 \begin{equation}
 \frac{\partial \rho(x,t)}{\partial t}=-\frac{\partial}{\partial x} \left[ \mu(x) \rho(x,t) \right] - (\gamma(x)+\kappa) \rho(x,t) +\langle \gamma(x) \rangle (t) \delta(x).
 \label{equation:master_gen}
 \end{equation}

In this continuous limit the last term is again the feeding at $x=0$ imposed by the Dirac delta function $\delta (x)$. This term allows to preserve the normalization of $\rho(x,t)$. The mean value of the reset rate ($\langle \gamma \rangle$) is given as:
 \begin{equation}
 \langle \gamma(x) \rangle (t) = \int_{\{x\}} (\gamma(x)+\kappa) \rho(x,t) dx
 \label{equation:feeding}
 \end{equation}
 
\noindent
In the stationary limit 
\begin{equation}
 \frac{\partial \rho(x,t)}{\partial t}=0,
 \end{equation}
the evolution equation for the probability density of having a tree size $x$ described by Equation~\ref{equation:master_gen} has a compact analytical solution that depends only on the form of the chosen growth and reset rates~\cite{LGGR,Biro-Neda}

\newcommand{\eadx}[1]{{\rm e}^{#1} }
 
\begin{equation}
\rho_s(x) \: = \: \frac{C}{\mu(x)} \, \eadx{-\int_{\{x\}}\frac{(\gamma(u)+\kappa)}{\mu(u)}du},
\label{stat-distr}
\end{equation}
with $C$ a normalization constant. 

\noindent
Based on the form of the $\mu(x)$ growth- and $\gamma(x)$ reset rates, the LGGR model is able to reproduce stationary probability distributions, $\rho_s(x,t)$, that are frequently encountered in complex systems~\cite{Biro-Neda,ZNeda1,ZNeda2,LGGR}.  
 
\section*{Tree-size distribution in the LGGR approximation}

We apply now the LGGR modeling framework for describing the dynamics of tree-size distribution. There are three main processes that drive this dynamics: a monotonic growth, the possibility of a reset (natural mortality, exploitation) and a multiplicative change in the number of trees belonging to one species. These stochastic processes are mathematically 
quantified by the $\mu(x)$ growth-rate, the $\gamma(x)$ reset-rate and $\kappa$ dilution rate. Once the needed kernel functions are realistically defined, the dynamics given by the LGGR model should yield the time evolution of the tree-size distribution function. In a general study of the LGGR dynamics it was previously shown~\cite{comments-on-biro-neda}, that apart of some pathologic cases, such systems are indeed converging to the stationary distribution. Depending on the starting condition, the mean of the distribution might converge slowly to a stationary value, however the distribution of $x/\langle x \rangle$ converges quickly to a stationary distribution. 
Given that the considered ecosystems (semi-natural forest and wood-pastures) are determined largely by mature trees, we can assume that the DBH distributions that we see in the forest and wood-pasture correspond to the stationary distribution. Definitely this is not the case for the plantations. Interestingly however, even in this clearly non-stationary case, their size-distribution during the growth process can be rescaled if we normalize the sizes to the mean value. This is what we see in Figure~\ref{fig:distribs-exp}b for the plantations: although the diameters are continuously increasing, the statistics in $x/\langle x \rangle$ is practically unchanged for a plantation that is 10 or 15 years old. This scaling, suggests that the growth speed of the trees have to increase as a function of the tree diameter, i.e. larger trees have to grow quicker. 

For choosing the right functional form for the growth and reset rates we take into account empirical knowledge for the tree life cycle, diversity dynamics in natural forest environments, previous experimental observations on such processes, and aim for a mathematical simplicity that allows compact analytical results. In contrast with the modeling methodology used by ecologists and biologists, we follow here a physicists approach for such complex systems, using a small number of model parameters and by simple, yet realistic, assumptions we aim to describe the main elements and universal features in the observed statistics. The confirmation of our model will not focus thus on the statistical goodness of the fit as it was done in the work of Lima \cite{de_Lima_2015} for example, but rather on the desire to understand  by a simple analytical model the dynamical mechanism leading to the universal form of the tree-size distribution in mature forest ecosystems. 

\subsection*{Growth rate}

Both our measurement data on the poplar tree plantations and the data available in the literature~\cite{growth_rate_afrika,Increment-Models,Volume-increment,Growth-redpine,South-Korea,volume-biomass,volume-dbh} supports the assumption that the growth rate ($\mu(x)$) of deciduous trees monotonically increases with the tree diameter. Even without a reset process this increase can not go on indefinitely, therefore for large trees it has to saturate. A mathematical form that can accommodate such a growth rate is:
\begin{equation}
 \mu(x) = d_{1}\frac{x}{x+b}, \quad b \geq 0
 \label{eq:growth-rate-theory}
 \end{equation}

The specific functional form, Equation \ref{eq:growth-rate-theory}, for the growth rate was taken by aiming to mathematical simplicity, however it's form and the involved $b$ parameter value is consistent with all  experimental data. More information for justifying this form and parameters is given in the \hyperref[sec:Discussion]{Discussion} section.

\subsection*{Reset rate}
Unlike, the growth rate, the reset rate is much more difficult to measure experimentally (it can be caused either by natural mortality or forest exploitation). The available data does not reflect the reset rate ($\gamma(x)$) itself, it yields instead the probability that a dead tree with a given diameter exists in an ecosystem~\cite{Urban-tree-mortality, tree-removal, tree-harvest, eu-forest-management}. In the framework of our modeling, this quantity is proportional to the product of the reset rate and probability density function, $~ \gamma(x) \cdot \rho(x)$. Similarly with the increasing growth rate as a function of tree sizes, assuming an increasing reset rate would be natural. Further evidences for this assumption is given in the next section. One would also expect that the reset rate is also converging to a constant value for very large trees. Deriving a reasonable kernel function for the reset rate is however more complicated. 

{\em First}, in all tree census data there is an $x_{min}$ minimal diameter under which trees do not enter in the statistics both for the dead and living trees. This means that from the viewpoint of the detected dynamics the reset should be negative (trees are just entering in the statistics) for $x<x_{min}$.  These facts, are all in agreement with a reset rate in the form:
\begin{equation}
\gamma(x)=f_1 \frac{x-r}{x+g}  \quad r \geq 0; \quad g>0 
\label{eq:gamma-original}
\end{equation}
Here $r$, $g$ and $f_1$, are positive constants. In  further calculations we will assume $g=b$, reducing the number of model parameters and making the mathematics simpler. 

{\em Second}, it is known that tree diversity increases in a deciduous mature forest. This means, that whenever a tree is dying, its place can be overtaken by an individual from another species. In the case of Oaks, for example, the establishemnt of young Oaks to replace the mature Oak trees in forests is hambered by improper light conditions. In such systems the likelihood for other, shade tolerant trees to replace the Oak is high. In the case of Beech and Hornbeam, both species tolerate and regenerate in shade - in these cases the replacement of old individuals can happen by same or different species (T. Hartel personal observation across Transylvanian deciduous forests). In both cases, the removal of mature trees represent a diversification of the forest stand for the semi-natural forest. Since the diversity is increasing, this effect will lead to a multiplicative decrease in the number of individuals for a species, which is equivalent (as we have shown in the previous section) with a $\kappa<0$ state independent reset term. Taking all these effects into account, we propose thus that the reset rate should be taken in the form
\begin{equation}
\gamma(x)=f_1 \frac{x-r}{x+b}+\kappa \equiv d_2\frac{x-c}{x+b},
\label{eq:reset-rate-final}
\end{equation}
with: 
\begin{equation}
c =\frac{f_1r-b\kappa}{f_1+\kappa} > r >0,
\end{equation}
\begin{equation}
d_2=(f_1+\kappa)<f_1; \quad \text{and} \quad d_2 > 0.
\label{eq:reset-rate-theory}
\end{equation}
Using data for the size-distribution of dead trees in several mature deciduous forests we can also verify whether the form of the proposed reset rate is a reasonable hypothesis. If we denote by $\rho_s(x)$ the stationary limit of the probability density for the DBH of the trees, the size distribution of the dead trees should follow the $\rho_s(x) \cdot \gamma(x)$ distribution with $\gamma(x)$ given by Equation \ref{eq:reset-rate-final}. This will be checked on the available data in the \hyperref[sec:Discussion]{Discussion} section, after we have derived the form of $\rho_s(x)$. Concerning the three investigated ecosystems with the applied cultivation approaches, the main triggering conditions for the tree mortality (reset) are summarized in Table~\ref{table:reset-in-systems}.

\begin{table}[!ht]
\centering
\setlength{\tabcolsep}{2pt} % Adjust inter-column spacing
\begin{tabular}{|p{2.2cm}|p{6.9cm}|p{6.9cm}|}
%\begin{tabular}{|p{0.14\textwidth}|p{0.43\textwidth}|p{0.43\textwidth}|}
\hline
{\bf System} & {\bf Management control} & {\bf Main drivers of tree mortality (interpretable as a contribution to  reset in our modelling)} \\ \hline
\parbox{2cm}{Mature forest with high natural values} & \parbox{6.8cm}{Weak, reduced to an initial oak plantation in the first part of the 1900s. The subsequent increase of the abundance of hornbeam, beech, and other tree species happened naturally. Natural regeneration and the accumulation of dead trees in the forest is accentuated. While timber exploitation happens (the Oak and Beech being valued), this is never at large scale, only at parcels of cca 1-3 hectares and when the trees have cca 90-120 years. Grazing is prohibited by law~\cite{Hartel_2013}.} \vspace{6pt} & \parbox{6.8cm}{Mostly inter- and intraspecific competition for light. In a lesser extent extreme meteorological conditions, pest outbreaks, fire and illegal cutting~\cite{Hartel_2013}.} \\ \hline
\parbox{2cm}{Ancient wood-pasture} & \parbox{6.8cm}{Weak, represented by traditional grazing with sheep, cattle, buffalo and other livestock as well as scrub clearance in the central parts of the pasture. Tree regeneration happens in pulses via associative resistance~\cite{Hartel_2013}. The oldest trees in such a system have over 300 years.} & \parbox{6.8cm}{Mostly extreme weather conditions (strong winds, lightening, and recently increasing drought)\\ weakening or damaging individual mature trees \\ which will be subsequently removed with formal\\ permit. Illegal fires set by shepherds can be also\\ a cause of mortality for old trees. Grazing prohibits tree regeneration in areas without shrubs. In a lesser extent competition and pests or diseases~\cite{Hartel_2013}} \vspace{6pt} \\ \hline
\parbox{2cm}{Plantation \\ under strong \\ management} & \parbox{6.8cm}{In case of our two plantations no direct human intervention has happened since the establishment.} 
 & \parbox{6.8cm}{The intraspecific (or even intraclonal) competion can be significant. As clone origin, the trees are almost identical genetically. Abiotic factors (wind/storms/snow) caused some level of disturbances.} \vspace{6pt} \\ \hline
\end{tabular}
\caption{Type of management control and main causes of tree mortality in the considered woodland areas.}
\label{table:reset-in-systems}
\end{table}

\subsection*{Stationary size distribution}

Once we accept the form given by Equations \ref{eq:growth-rate-theory} and \ref{eq:reset-rate-final} for the growth and reset rates, respectively, it is straightforward to compute the stationary probability density, $\rho_s(x)$. Since the $\kappa$ value has been now incorporated in the $\gamma(x)$  reset rate (Equation \ref{eq:reset-rate-final}), according to Equation \ref{stat-distr} we get
\begin{equation}
\rho_s(x) \: = \: \frac{\mu(0) \rho_s(0)}{\mu(x)} \, \eadx{-\int_{\{x\}} \frac{\gamma(u)}{\mu(u)}du} = C x^{d \,c-1} (x+b)\,
\eadx{-d \,x},
\end{equation} 
where $d=d_2/d_1$ and $C$ is a normalization constant. If the distribution is defined on the $x \in [0,\infty)$ interval, the normalization constant becomes:
\begin{equation}
C=\frac{d^{c\, d}}{(b+c)\Gamma[c\,d]}.
\label{eq:normalization}
\end{equation}
The first moment of the distribution (average) is also analytical:
\begin{equation}
\langle x \rangle =c \left ( 1+\frac{1}{(b+c)d} \right ).
\end{equation}
We write now the distribution function for the $y=x/\langle x \rangle$ tree-sizes normalized relative to the mean value:
\begin{equation}
\rho_s(y)=\frac{d^{c\, d}}{(b+c)\Gamma[c\,d]} \langle x \rangle ^{d c}
\eadx{-d \langle x \rangle y} y^{dc-1} (y \langle x \rangle +b).
\end{equation}
Assuming  that $\langle x \rangle=1$, it results
\begin{equation}
b=\frac{c}{(1-c)d}-c,
\end{equation}
therefore the probability density function will have only two-parameters to fit the experimental results for $y=x/\langle x \rangle$:
\begin{equation}
\rho_s(y)=\frac{d^{c\, d}}{(\frac{c}{(1-c)d})\Gamma[c\,d]} 
\eadx{-d  y} y^{dc-1} \left(y+\frac{c}{(1-c)d}-c\right).
\label{eq:density-function-final}
\end{equation}
As Figure \ref{fig:distribs-exp}a shows, the probability density for the distribution of $x/\langle x \rangle$ on forests and wood-pastures collapse, and it can be well approximated by the form given in Equation \ref{eq:density-function-final}, with parameters $c=0.8$ and $d=4.9$, leading to $b=0.016$.

\section*{Discussion}
\label{sec:Discussion}

We discuss now our main findings and comment on the model parameters that lead to qualitatively good description of tree diameter distribution of deciduous tree species in mature forest environments.

{\bf Growth, reset and diversification.} We have shown that the Gamma distribution describes well the tree-size distribution in both forest and  wood-pasture environments. Interestingly, the probability density functions of $DBH$ for different species, and different environments collapse, when the distribution of $DBH/\langle DBH \rangle$ is constructed. This intriguing
universality is captured by our model if we assume the same $c$ and $d$ parameters for all species and for the different environments (forest and wood-pastures). This means that in the tree census one has to consider the same lower limits for recording a tree, the same dilution rate,  $\kappa$, due to diversification and the ratio of the reset and growth rates should be similar for the same $y=x/\langle x \rangle$ relative diameter values. These are all in agreement with the fact that the considered three deciduous genera dominate quite equally these forest environments and they are ecologically equally fit. 

The LGGR modeling framework was able to reproduce the observed distribution, assuming three competing processes that affect the tree sizes and tree numbers in a mature woodland. 

\begin{figure}[ht]
\centering
\includegraphics[width=0.8\textwidth]{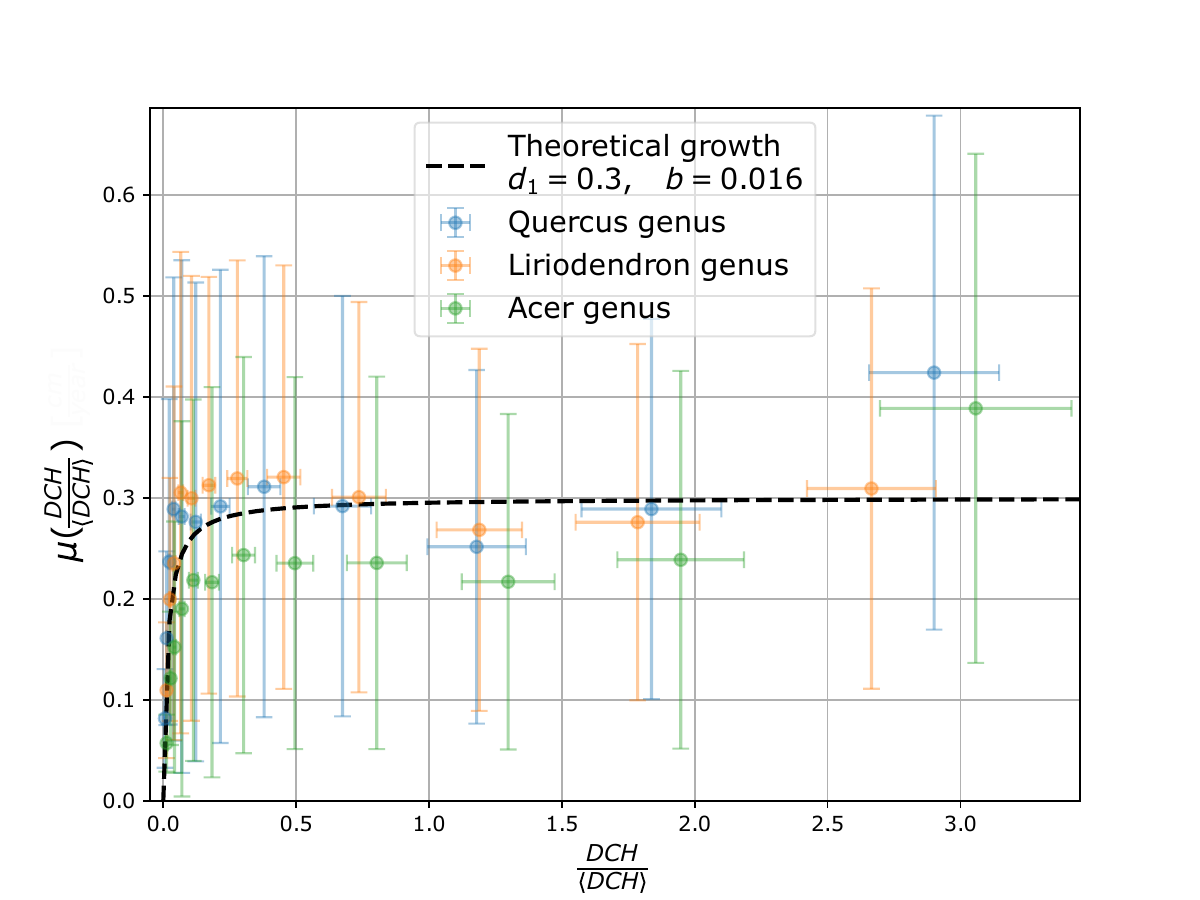}
\caption{
Growth rate determined from the width of tree rings. The figure illustrates the width of tree rings as a function of stem diameter at one meter above the ground for three tree genera as indicated in the legend. The trend illustrated by the dashed line is given by Equation \ref{eq:growth-rate-theory} with parameters indicated in the figure.}
\label{fig:growth-rate}
\end{figure}

The first process is a {\em monotonic growth}, which was assumed to increase with tree size and saturate in the limit of large diameters. The growth rate given in Equation \ref{eq:growth-rate-theory} is supported by the data provided by the United States National Park Service (\verb|NPS|)~\cite{NCRN:NPS,NCRN_method}, where we have identified the annual growth rate from the diameter of the tree rings. For three tree genera (Quercus, Liriodendron and Acer) on Figure \ref{fig:growth-rate} we plot the averaged annual growth rate as a function of  $DCH/\langle DCH\rangle$ ($DCH$ stands for the Diameter at Core Height, the core height is approximately 1 m above the ground).
The number of trees by genera that were considered for computing these growth rates were: {\it Quercus genus} 545 trees ({\it (Quercus Alba, Quercus Rubra, Quercus Montana} species) ; {\it Liriodendron genus} 210 trees ({\it Liriodendron Tulipifera species}); {\it Acer genus} 64 trees ({\it Acer Negundo, Acer Rubrum, Acer Saccharinum} species). On the same figure, we also indicate the trend that is given by the kernel function for the growth rate, Equation \ref{eq:growth-rate-theory}, with a parameter set that gives a reasonable description of the data.

\begin{figure}[ht]
\centering
\includegraphics[width=0.8\textwidth]{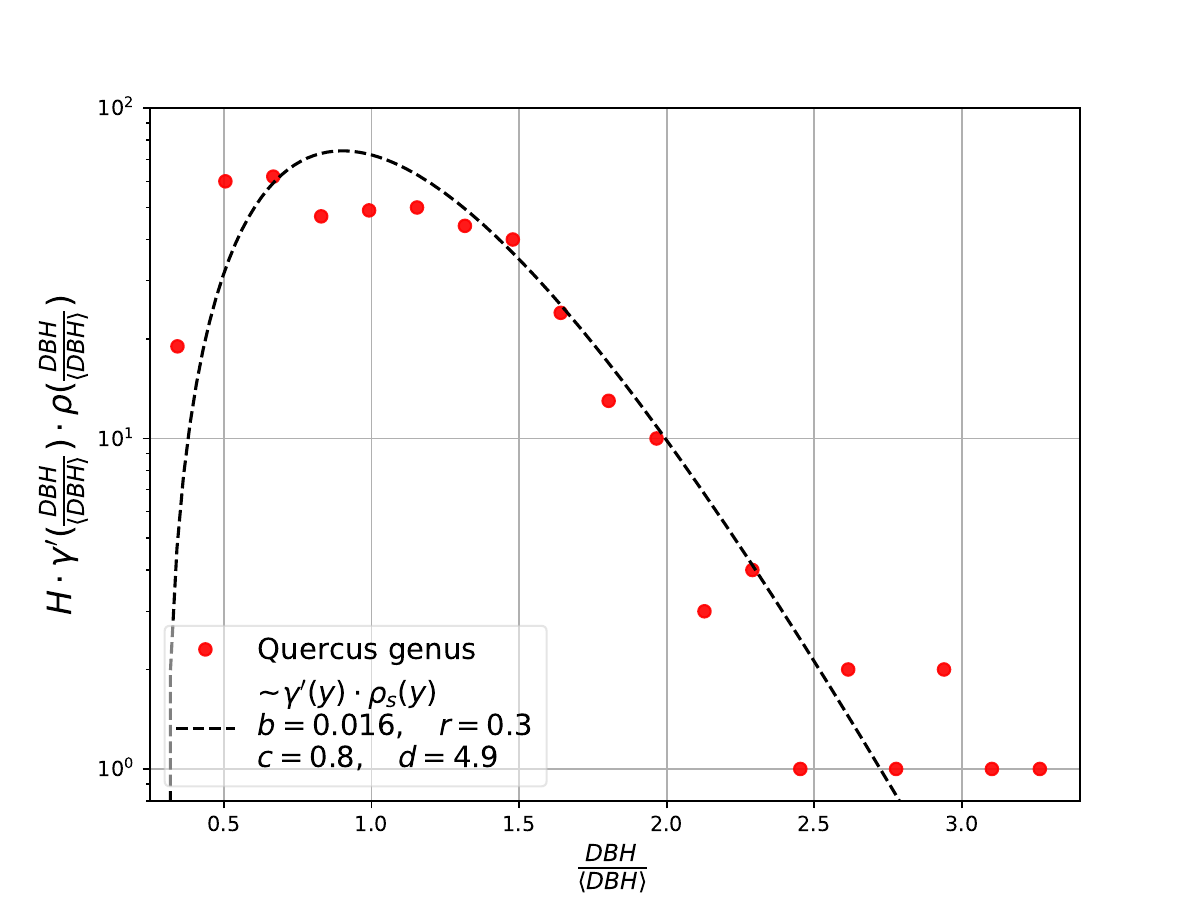}
\caption{Consistency between the dead trees size distribution, the considered reset rate and fitted tree-size distribution. Histogram of the size distribution (sizes normalized to the mean) of dead Quercus trees from censused forests plotted together with the fit given by $H \cdot \rho_s(y) \gamma'(y)$ with $r=0.3$ and parameters estimated from the probability density function. $H$ is a constant needed to fit the experimental histogram.}
\label{fig:reset-rate}
\end{figure}

A second process which complements this growth and allows for developing a stationary distribution is {\em tree mortality}, captured by our reset rate. In order to derive a mathematical form for this rate, we considered a process where there is a lower $r$ limit for detecting a new or dead tree in the census (trees below this size are not measured). According to this methodology below the $r$ size, trees are appearing in the statistics, a process that can be taken into account with a negative reset rate. We assumed also that tree mortality rate should increase in a forest environment with tree sizes due to both endogenous and exogenous effects. Similarly with the growth rate, this should saturate to a constant value for large trees. A mathematically simple reset rate that could reproduce these features was proposed in the form given by Equation \ref{eq:gamma-original}. The trend given by this form is also supported by the literature~\cite{Urban-tree-mortality, tree-removal, tree-harvest, eu-forest-management}. As we have emphasized in the previous section and will show here, this reset rate together with the proposed form of the probability density function (Equation \ref{eq:density-function-final}) leads to results that are in agreement with observations.  For testing this reset rate we can use again the data from \verb|NPS|~\cite{NCRN:NPS,NCRN_method} for dead trees diameter, which should be fitted as $\rho_s(y) \cdot \gamma'(y)$ with the $\gamma'(y)=f_1\cdot (y-r)/(y+b)$ form of the reset. The data provided by the United States National Park Service, contains the diameter of dead trees within a number of 320 plots from 10 national parks from the USA. For consistency, and for putting together several data from different forests, the trees' diameter is normalized to the mean value of tree  diameters in the forest (taking now only the living trees). Considering the Quercus genus, the data for the histogram of the dead trees is plotted in Figure \ref{fig:reset-rate}. The dashed line indicates a fit based on Equation \ref{eq:density-function-final} with the parameters $c=0.8$, $b=0.016$ and $d=4.9$ for the experimentally observed probability density and $r=0.3$ in the $\gamma'(y)$ reset rate.

Finally, in order to explain the large $c=0.8$ value in the final form of the reset rate (Equation \ref{eq:reset-rate-final}), which is necessary for a reasonably good fit of the diameter distributions, we had to assume another reset-like process, due to the {\em diversification process} implying competitive exclusion of certain species by other species. As it was shown in the general discussion (\hyperref[sec:LGGR-general]{The LGGR modeling framework}), a multiplicative growth or dilution in the total tree number belonging to a species is equivalent with a reset term in the master equation for the probability density function. Diversification is a known process in mature deciduous forest environments~\cite{Yeom-Kim:2011, Dupre:2002}, therefore our model had to consider this, and without this process one would not be able to explain the large $c$ value ($c=0.8$) in the fit of the probability density function from Equation \ref{eq:density-function-final}.

{\bf Consistency in the model parameters.} We crosscheck, whether the fit parameters for the experimentally observed probability density function is in agreement with the data that we have on growth and reset processes. 

\begin{figure}[ht]
\centering
\includegraphics[width=0.8\textwidth]{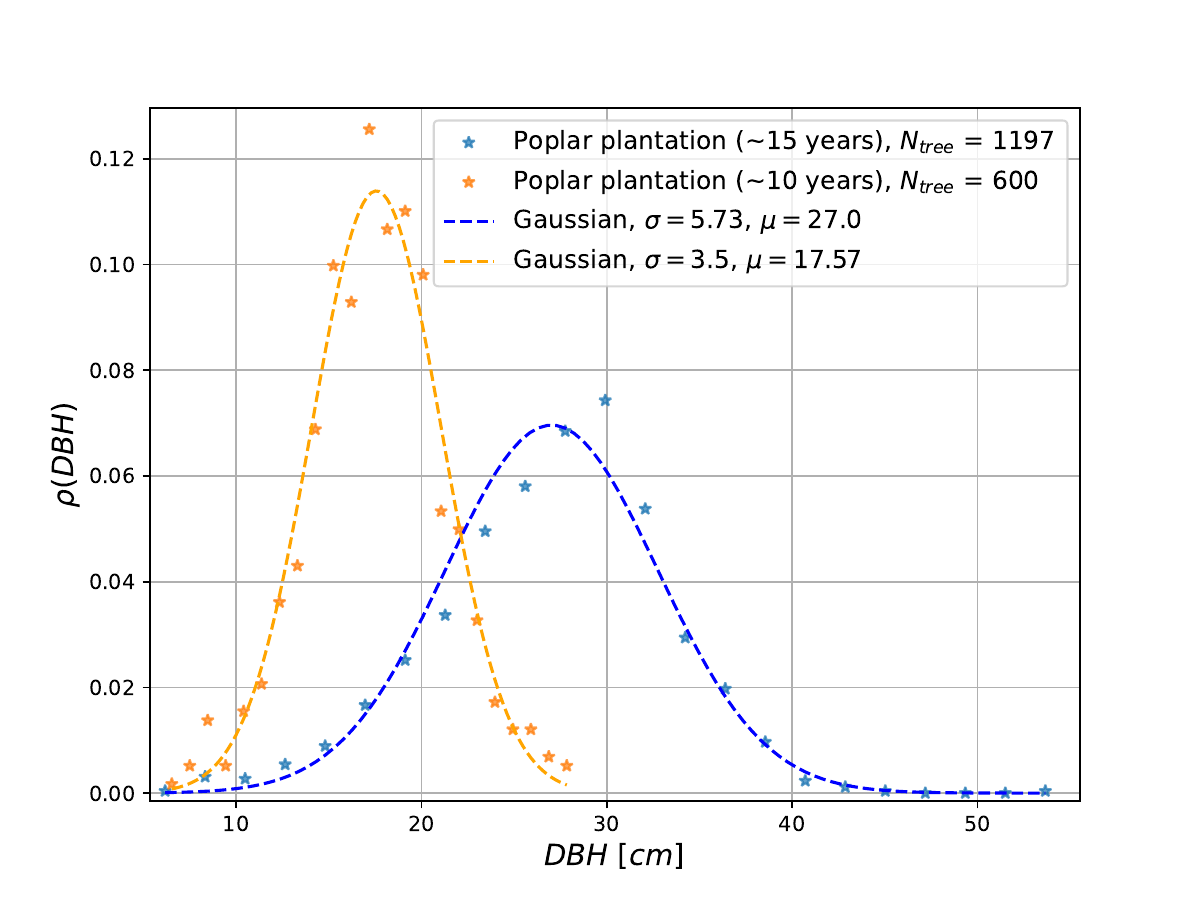}
\caption{Evolution of the size-distribution in poplar tree plantations. Tree size distribution (both fitted with a Gaussian) in a 10 and 15 year old poplar tree plantations with very similar ecological background.}
 \label{fig:plantation-not-rescaled}
\end{figure}

For the estimated growth-rate, Figure \ref{fig:growth-rate} shows that the parameter $b=0.016$ taken from the fitted probability density function is appropriate for a reasonable fit.  Concerning the monotonically growing nature of the growth rate as a function of tree sizes, a qualitative evidence is also our measurement in poplar tree plantations, where the standard deviation of the tree-size distribution increases as the mean size increases. Plotting (Figure \ref{fig:plantation-not-rescaled}) the measured size-distribution in the 10 and 15 years old plantations, without normalizing the tree sizes to the mean value, clearly indicates this trend. This is in agreement with only a monotonically increasing $\mu(x)$ growth rate.

Also in agreement with our 
prediction and imposed restrictions we find that the  best $r$ parameter value for fitting the reset data, satisfies the $r< c$  condition. The other parameters used for the fit shown in Figure \ref{fig:reset-rate} are the same as the ones used to fit the probability density function of tree-size distributions in Figure \ref{fig:distribs-exp}a.
Because we have no information on when these trees dried out, no direct values of the rates can be estimated and as a consequence one cannot determine the $f_1$ parameter that would allow estimation of the $\kappa$ parameter as well.
 
\begin{figure}[ht]
\centering
\includegraphics[width=0.7\textwidth]{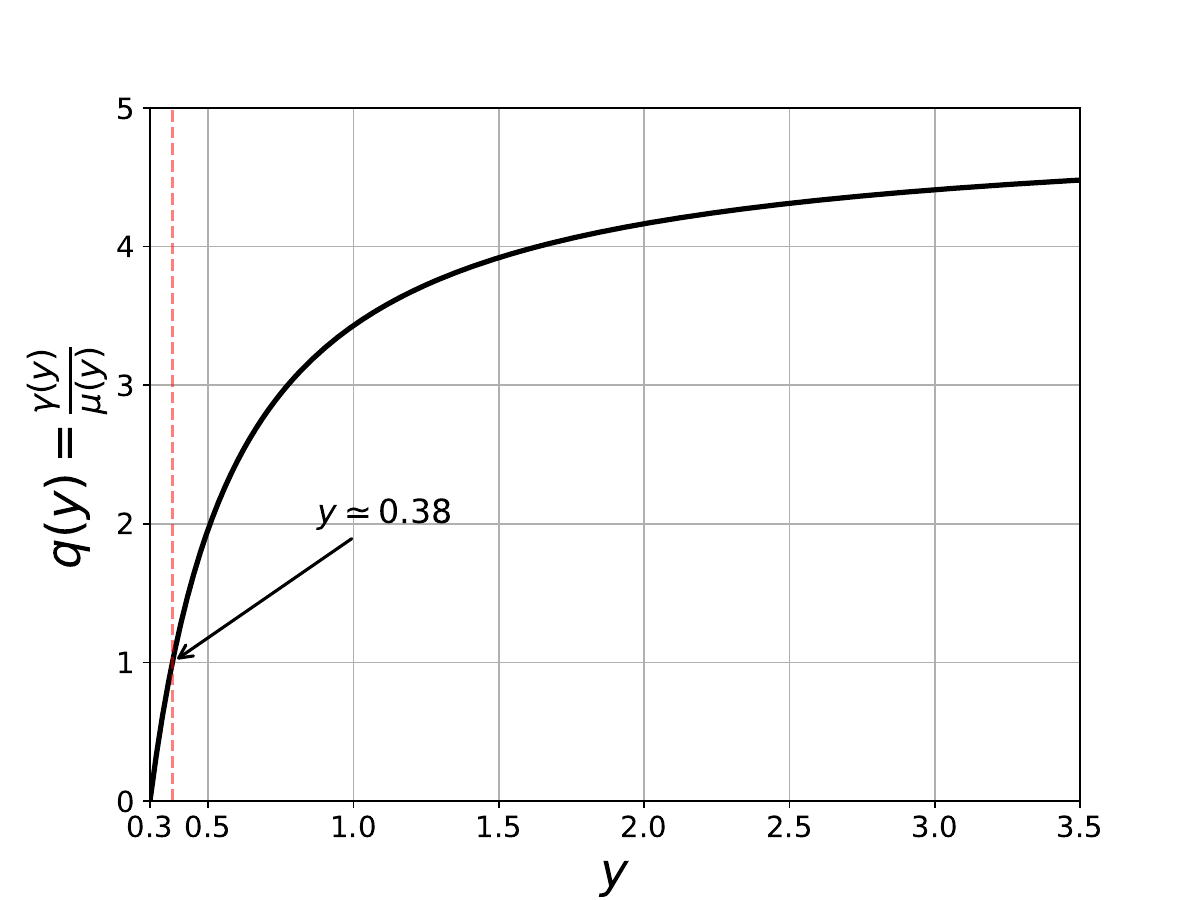}
\caption{Reset over growth rates probability for the genus Quercus as a function of the trees $y=x/\langle x \rangle$ relative size. The trend of $q(y)=\gamma'(y)/\mu(y)=d(y-r)/y$ for $r=0.3$ and $d=4.9$. }
 \label{fig:trend-q}
\end{figure}

Accepting the $r=0.3$ parameter from the fit in Figure \ref{fig:reset-rate}, we can also predict the reset rate over growth rate ratio ($q=\gamma'(y)/\mu(y)$) as a function of tree diameters (all sizes taken relative to the mean value). We get:
\begin{equation}
q=d\, \frac{y-r}{y}
\end{equation}
Using the $d$ value obtained from the fitted probability density function, $d=4.9$, and the $r=0.3$ value the $q(y)$ trend is plotted in Figure \ref{fig:trend-q}.

From this figure we learn, that the ratio $q$ is monotonically increasing as function of tree sizes and for trees over $y>0.38$ the reset process is more probable than growth. This intuitively explains why despite the monotonically increasing growth rate the forest does not get filled up by very large trees.

{\bf Rigorous statistical analyses versus an elegant modeling framework.} In fitting the experimental data and analyzing the goodness of the fit our aims were quite modest and we followed basically a physicist modeling methodology. Our main interest focused on unveiling some interesting universality and to show the visually acceptable collapse of the renormalized data. The best fit parameters were estimated by using just a visual comparison with the experimental data, however we insisted on the model parameters consistency for all the analyzed data. Although in such a view it does not make much sense to make a statistical goodness analyses of the regressions, neither to interpret the deviations from the experimental data in order to emphasize the appropriateness of the designed model, we still listed the values of the coefficients of determination ($R^2$) in Table \ref{table:determination-natural}. Instead of a  rigorous quantitative modeling with many unknown parameters, we opted for a simple analytically solvable model with basically two free parameters. One should consider the mathematical forms for the $\mu(x)$ growth rate and $\gamma(x)$ reset rates in the evolutionary master equation, as mathematically convenient first order approximations that satisfy some imposed restrictions.
Definitely, one can come up with other, more accurate forms for these kernel functions, describing better the experimental data. The drawback with such attempt will be the more complicated form for the stationary probability density and the inevitable increase in the number of model parameters. The available data itself 
was barely enough to construct the qualitative form of the probability density functions, and as it is visible on Figure \ref{fig:distribs-exp} it has large deviations from a smooth trend. In such conditions the best one can do is to offer a visually good fit for the data with a consistent theoretical description of the underlying processes.  

\begin{table}[!ht]
\centering
\begin{tabular}{|l|l|l|l|}
\hline
{\bf Species} & {\bf Oak} & {\bf Beech} & {\bf Hornbeam}  \\ \hline
Forest & 0.91 & 0.81 & 0.97 \\ \hline
Wood-pasture & 0.8 & 0.46 & 0.86 \\ \hline
\end{tabular}
\caption{Coefficient of determination ($R^2$) for the Gamma fit of the the tree-size distribution in 
natural woodlands}
\label{table:determination-natural}
\end{table}

\section*{Conclusion}

Tree size diversity patterns in natural deciduous forest and wood-pasture environments is a complex problem, where new data and simple realistic mathematical models are needed for its better understanding. It has been conjectured that the diameter distribution of trees belonging to given deciduous species follows a Gamma distribution in a mature natural forest. Here we brought new evidences supporting this hypothesis, considering new exhaustive measurement data for three tree genera in two different environments: mature semi-natural forest and wood-pastures located in Sibiu county, Romania. 

Apart of the generality for the Gamma distribution, our data suggests an intriguing statistical universality: rescaling the tree diameters with the average tree diameter for that species in the given forest, all the data collapsed on the very same distribution. Seemingly we deal thus with some interesting stylized facts in tree-size diversity patterns for deciduous temperate climate forests, allowing also a useful rescaling among different species and different natural ecosystems. Data collected on relatively young (up to 15 years) tree plantations, reveal different diversity patterns. These plantations clearly did not reach maturity and a stationary state, therefore the difference relative to what is observed in the mature forest environments should not be a surprise at all. This findings suggests that the Gamma type fit for the tree-size distribution can be used as a simple proxy to infer woodland naturalness and maturity.

In order to understand theoretically the tree-size distribution in natural forest environments the main processes that govern the evolution of the tree ensemble has to be considered: growth, mortality and a general diversification of tree species. The easiest way to elaborate a model that is able to predict a stationary tree-size distribution is to incorporate these probabilistic processes in an evolutionary master equation. This has been done here, in the framework of the previously introduced LGGR model. By considering mathematically simple, but still realistic forms for the growth and reset processes, supported also by experimental data, the stationary distribution provided by the LGGR model reproduced successfully the experimental results. In the modeling process we followed the physicists way of thinking. We aimed to obtain compact analytical results that describes visually well the experimental data by using a very small number of free parameters in the model. By doing this we concentrated less on the statistical goodness of the provided fit and insisted more on mathematical simplicity and the usefulness of analytical results in a compact mathematical form. Taking into account that the experimental data used for testing the growth and reset rate is quite poor and their sources are diverse, we consider that  this approach is more fruitful for understanding the experimentally observed universal shape of tree-size distributions. 

Naturally, in order to get further confidence in the proposed model, new and good quality data should still be gathered. It would be interesting to test in the very same forest and wood-pasture environment the growth and reset dynamics of the considered tree species. Within the same forest it would be also interesting to gather
quantitative data on the diversification process for the tree species. For doing this however, controlled  tree census measurements have to be planned and continuously repeated.

\bibliography{tree_bib}

\section*{Acknowledgments}

Work supported by the UEFISCDI PN-III-P4-ID-PCE-2020-0647 research grant. The work of Sz.K. and M.J. is also supported by the Collegium Talentum Program of Hungary. We are thankful to E. Gabnai for helping us to choose the hybrid poplar plantations as well to Zs. N\'{e}da, M. Paulin and Cs. G\'{a}sp\'{a}r for their help in the field work.

\section*{Author contributions statement}

Z.N. conceived the model and designed the study,  Sz.K. made the data analyses, unified the experimental data and constructed the figures, M.J. participated in data analyzes and model validation, T.H. and Gy.Cs. provided the experimental data and consultancy from the biological perspective. First draft of the manuscript by  Z.N and Sz.K.  All authors reviewed the manuscript. 

\section*{Competing interests}

The authors declare no competing interests.

\section*{Data availability}

The data collected by the authors (summarized in Table \ref{tab:data-all}) are freely available for download from:~\cite{data:DBH:data}.
The data used for plotting Figures \ref{fig:growth-rate} and \ref{fig:reset-rate} are from the mentioned sources, and can be obtained by request. 

\section*{Compliance statement}

Our research, involving non-invasive measurements, fully adheres to the regulations of the International Union for Conservation of Nature (IUCN) Policy on Species at Risk of Extinction and the Convention on the Trade in Endangered Species of Wild Fauna and Flora (CITES) to ensure the ethical treatment and protection of endangered plant species.

\end{document}